%% file: 2019OA.tex
\documentclass[aps,prl,twocolumn,showpacs,superscriptaddress,nofootinbib,floatfix,amsmath,amssymb]{revtex4-1}

\usepackage{amsmath}   
\usepackage{graphicx}  
\graphicspath{{figures/}}
\usepackage{xspace}
\usepackage{units}
\usepackage{accents}
\usepackage{scalerel}
\usepackage{color}
\usepackage{bm}
\usepackage{subfigure}
\usepackage{alphalph}

\usepackage{hyperref}

\newlength{\heightnu}
\AtBeginDocument{\settoheight{\heightnu}{$\nu$}}

\newcommand{\sizecheck}{0} 
\newcommand{\PRLsupp}{1}   

\newcommand{\sinsqthetamu}{$\sin^2\theta_{23}$}
\newcommand{\sinsqthetamub}{$\sin^2\bar\theta_{23}$}
\newcommand{\Deltamu}{$\Delta m^2_{32}$}
\newcommand{\Deltamub}{$\overline{\Delta m^2_{32}}$}

\newcommand{\Dmunull}{${\Delta}m^{2}$}
\newcommand{\Dmunullb}{$\Delta\overline{m}^{2}$}
\newcommand{\Dmueff}{${\Delta}m^{2}$}
\newcommand{\Dmueffb}{${\overline{\Delta m^{2}}}$}
\newcommand{\evmass}{$\times 10^{-3} \text{eV}^{2}/c^{4}$ }
\ifnum\PRLsupp=0
  
\else
  
\fi

\newif\ifpdf
\ifx\pdfoutput\undefined
   \pdffalse
\else
   \pdfoutput=1
   \pdftrue
\fi
\ifpdf
   \usepackage{graphicx}
   \usepackage{epstopdf}
\else
   \usepackage{graphicx}
\fi

\begin{document}

\title{
    T2K measurements of muon neutrino and antineutrino disappearance using $3.13\times 10^{21}$ protons on target
}

\input{AuthorList_oa2019numunumubar_191020}

\date{\today}
\begin{abstract}
\input{abstract.tex}

\end{abstract}

\maketitle



\clearpage

\input{bodytext.tex}

\ifnum\sizecheck=0
  \input{acknowledgments.tex}
\fi

\bibliographystyle{ieeetr} 
\bibliography{2019OA}

\end{document}

%% file: AuthorList_oa2019numunumubar_191020.tex

\newcommand{\INSTHD}{\affiliation{University Autonoma Madrid, Department of Theoretical Physics, 28049 Madrid, Spain}}
\newcommand{\INSTEE}{\affiliation{University of Bern, Albert Einstein Center for Fundamental Physics, Laboratory for High Energy Physics (LHEP), Bern, Switzerland}}
\newcommand{\INSTFE}{\affiliation{Boston University, Department of Physics, Boston, Massachusetts, U.S.A.}}
\newcommand{\INSTGA}{\affiliation{University of California, Irvine, Department of Physics and Astronomy, Irvine, California, U.S.A.}}
\newcommand{\INSTI}{\affiliation{IRFU, CEA, Universit\'e Paris-Saclay, F-91191 Gif-sur-Yvette, France}}
\newcommand{\INSTGB}{\affiliation{University of Colorado at Boulder, Department of Physics, Boulder, Colorado, U.S.A.}}
\newcommand{\INSTFG}{\affiliation{Colorado State University, Department of Physics, Fort Collins, Colorado, U.S.A.}}
\newcommand{\INSTFH}{\affiliation{Duke University, Department of Physics, Durham, North Carolina, U.S.A.}}
\newcommand{\INSTBA}{\affiliation{Ecole Polytechnique, IN2P3-CNRS, Laboratoire Leprince-Ringuet, Palaiseau, France }}
\newcommand{\INSTEF}{\affiliation{ETH Zurich, Institute for Particle Physics and Astrophysics, Zurich, Switzerland}}
\newcommand{\INSTIE}{\affiliation{CERN European Organization for Nuclear Research, CH-1211 Genève 23, Switzerland}}
\newcommand{\INSTEG}{\affiliation{University of Geneva, Section de Physique, DPNC, Geneva, Switzerland}}
\newcommand{\INSTHJ}{\affiliation{University of Glasgow, School of Physics and Astronomy, Glasgow, United Kingdom}}
\newcommand{\INSTDG}{\affiliation{H. Niewodniczanski Institute of Nuclear Physics PAN, Cracow, Poland}}
\newcommand{\INSTCB}{\affiliation{High Energy Accelerator Research Organization (KEK), Tsukuba, Ibaraki, Japan}}
\newcommand{\INSTIB}{\affiliation{University of Houston, Department of Physics, Houston, Texas, U.S.A.}}
\newcommand{\INSTED}{\affiliation{Institut de Fisica d'Altes Energies (IFAE), The Barcelona Institute of Science and Technology, Campus UAB, Bellaterra (Barcelona) Spain}}
\newcommand{\INSTEC}{\affiliation{IFIC (CSIC \& University of Valencia), Valencia, Spain}}
\newcommand{\INSTHH}{\affiliation{Institute For Interdisciplinary Research in Science and Education (IFIRSE), ICISE, Quy Nhon, Vietnam}}
\newcommand{\INSTEI}{\affiliation{Imperial College London, Department of Physics, London, United Kingdom}}
\newcommand{\INSTGF}{\affiliation{INFN Sezione di Bari and Universit\`a e Politecnico di Bari, Dipartimento Interuniversitario di Fisica, Bari, Italy}}
\newcommand{\INSTBE}{\affiliation{INFN Sezione di Napoli and Universit\`a di Napoli, Dipartimento di Fisica, Napoli, Italy}}
\newcommand{\INSTBF}{\affiliation{INFN Sezione di Padova and Universit\`a di Padova, Dipartimento di Fisica, Padova, Italy}}
\newcommand{\INSTBD}{\affiliation{INFN Sezione di Roma and Universit\`a di Roma ``La Sapienza'', Roma, Italy}}
\newcommand{\INSTEB}{\affiliation{Institute for Nuclear Research of the Russian Academy of Sciences, Moscow, Russia}}
\newcommand{\INSTHI}{\affiliation{International Centre of Physics, Institute of Physics (IOP), Vietnam Academy of Science and Technology (VAST), 10 Dao Tan, Ba Dinh, Hanoi, Vietnam}}
\newcommand{\INSTHA}{\affiliation{Kavli Institute for the Physics and Mathematics of the Universe (WPI), The University of Tokyo Institutes for Advanced Study, University of Tokyo, Kashiwa, Chiba, Japan}}
\newcommand{\INSTID}{\affiliation{Keio University, Department of Physics, Kanagawa, Japan}}
\newcommand{\INSTIF}{\affiliation{King's College London, Department of Physics, Strand, London WC2R 2LS, United Kingdom}}
\newcommand{\INSTCC}{\affiliation{Kobe University, Kobe, Japan}}
\newcommand{\INSTCD}{\affiliation{Kyoto University, Department of Physics, Kyoto, Japan}}
\newcommand{\INSTEJ}{\affiliation{Lancaster University, Physics Department, Lancaster, United Kingdom}}
\newcommand{\INSTFC}{\affiliation{University of Liverpool, Department of Physics, Liverpool, United Kingdom}}
\newcommand{\INSTFI}{\affiliation{Louisiana State University, Department of Physics and Astronomy, Baton Rouge, Louisiana, U.S.A.}}
\newcommand{\INSTHB}{\affiliation{Michigan State University, Department of Physics and Astronomy,  East Lansing, Michigan, U.S.A.}}
\newcommand{\INSTCE}{\affiliation{Miyagi University of Education, Department of Physics, Sendai, Japan}}
\newcommand{\INSTDF}{\affiliation{National Centre for Nuclear Research, Warsaw, Poland}}
\newcommand{\INSTFJ}{\affiliation{State University of New York at Stony Brook, Department of Physics and Astronomy, Stony Brook, New York, U.S.A.}}
\newcommand{\INSTGJ}{\affiliation{Okayama University, Department of Physics, Okayama, Japan}}
\newcommand{\INSTCF}{\affiliation{Osaka City University, Department of Physics, Osaka, Japan}}
\newcommand{\INSTGG}{\affiliation{Oxford University, Department of Physics, Oxford, United Kingdom}}
\newcommand{\INSTIC}{\affiliation{University of Pennsylvania, Department of Physics and Astronomy,  Philadelphia, PA, 19104, USA.}}
\newcommand{\INSTGC}{\affiliation{University of Pittsburgh, Department of Physics and Astronomy, Pittsburgh, Pennsylvania, U.S.A.}}
\newcommand{\INSTFA}{\affiliation{Queen Mary University of London, School of Physics and Astronomy, London, United Kingdom}}
\newcommand{\INSTGD}{\affiliation{University of Rochester, Department of Physics and Astronomy, Rochester, New York, U.S.A.}}
\newcommand{\INSTHC}{\affiliation{Royal Holloway University of London, Department of Physics, Egham, Surrey, United Kingdom}}
\newcommand{\INSTBC}{\affiliation{RWTH Aachen University, III. Physikalisches Institut, Aachen, Germany}}
\newcommand{\INSTFB}{\affiliation{University of Sheffield, Department of Physics and Astronomy, Sheffield, United Kingdom}}
\newcommand{\INSTDI}{\affiliation{University of Silesia, Institute of Physics, Katowice, Poland}}
\newcommand{\INSTBB}{\affiliation{Sorbonne Universit\'e, Universit\'e Paris Diderot, CNRS/IN2P3, Laboratoire de Physique Nucl\'eaire et de Hautes Energies (LPNHE), Paris, France}}
\newcommand{\INSTEH}{\affiliation{STFC, Rutherford Appleton Laboratory, Harwell Oxford,  and  Daresbury Laboratory, Warrington, United Kingdom}}
\newcommand{\INSTCH}{\affiliation{University of Tokyo, Department of Physics, Tokyo, Japan}}
\newcommand{\INSTBJ}{\affiliation{University of Tokyo, Institute for Cosmic Ray Research, Kamioka Observatory, Kamioka, Japan}}
\newcommand{\INSTCG}{\affiliation{University of Tokyo, Institute for Cosmic Ray Research, Research Center for Cosmic Neutrinos, Kashiwa, Japan}}
\newcommand{\INSTHF}{\affiliation{Tokyo Institute of Technology, Department of Physics, Tokyo, Japan}}
\newcommand{\INSTGI}{\affiliation{Tokyo Metropolitan University, Department of Physics, Tokyo, Japan}}
\newcommand{\INSTHG}{\affiliation{Tokyo University of Science, Faculty of Science and Technology, Department of Physics, Noda, Chiba, Japan}}
\newcommand{\INSTF}{\affiliation{University of Toronto, Department of Physics, Toronto, Ontario, Canada}}
\newcommand{\INSTB}{\affiliation{TRIUMF, Vancouver, British Columbia, Canada}}
\newcommand{\INSTDJ}{\affiliation{University of Warsaw, Faculty of Physics, Warsaw, Poland}}
\newcommand{\INSTDH}{\affiliation{Warsaw University of Technology, Institute of Radioelectronics and Multimedia Technology, Warsaw, Poland}}
\newcommand{\INSTFD}{\affiliation{University of Warwick, Department of Physics, Coventry, United Kingdom}}
\newcommand{\INSTGH}{\affiliation{University of Winnipeg, Department of Physics, Winnipeg, Manitoba, Canada}}
\newcommand{\INSTEA}{\affiliation{Wroclaw University, Faculty of Physics and Astronomy, Wroclaw, Poland}}
\newcommand{\INSTHE}{\affiliation{Yokohama National University, Department of Physics, Yokohama, Japan}}
\newcommand{\INSTH}{\affiliation{York University, Department of Physics and Astronomy, Toronto, Ontario, Canada}}

\INSTHD
\INSTEE
\INSTFE
\INSTGA
\INSTI
\INSTGB
\INSTFG
\INSTFH
\INSTBA
\INSTEF
\INSTIE
\INSTEG
\INSTHJ
\INSTDG
\INSTCB
\INSTIB
\INSTED
\INSTEC
\INSTHH
\INSTEI
\INSTGF
\INSTBE
\INSTBF
\INSTBD
\INSTEB
\INSTHI
\INSTHA
\INSTID
\INSTIF
\INSTCC
\INSTCD
\INSTEJ
\INSTFC
\INSTFI
\INSTHB
\INSTCE
\INSTDF
\INSTFJ
\INSTGJ
\INSTCF
\INSTGG
\INSTIC
\INSTGC
\INSTFA
\INSTGD
\INSTHC
\INSTBC
\INSTFB
\INSTDI
\INSTBB
\INSTEH
\INSTCH
\INSTBJ
\INSTCG
\INSTHF
\INSTGI
\INSTHG
\INSTF
\INSTB
\INSTDJ
\INSTDH
\INSTFD
\INSTGH
\INSTEA
\INSTHE
\INSTH

\author{K.\,Abe}\INSTBJ
\author{N.\,Akhlaq}\INSTFA
\author{R.\,Akutsu}\INSTHA
\author{A.\,Ali}\INSTCD
\author{C.\,Alt}\INSTEF
\author{C.\,Andreopoulos}\INSTEH\INSTFC
\author{M.\,Antonova}\INSTEC
\author{S.\,Aoki}\INSTCC
\author{T.\,Arihara}\INSTGI
\author{Y.\,Asada}\INSTHE
% ...

\collaboration{The T2K Collaboration}\noaffiliation

%% file: abstract.tex
 We report measurements by the T2K experiment of the parameters $\theta_{23}$ and $\Delta m^2_{32}$ which govern the disappearance of muon neutrinos and antineutrinos in the three-flavor PMNS neutrino oscillation model at T2K's neutrino energy and propagation distance. Utilizing the ability of the experiment to run with either a mainly neutrino or a mainly antineutrino beam, muon-like events from each beam mode are used to measure these parameters separately for neutrino and antineutrino oscillations. Data taken from $1.49 \times 10^{21}$ protons on target (POT) in neutrino mode and $1.64 \times 10^{21}$ POT in antineutrino mode are used. The best-fit values obtained by T2K were $\sin^2\left(\theta_{23}\right)=0.51^{+0.06}_{-0.07} \left(0.43^{+0.21}_{-0.05}\right)$ and $\Delta m^2_{32}=2.47^{+0.08}_{-0.09} \left(2.50^{+0.18}_{-0.13}\right)$\evmass for neutrinos (antineutrinos). No significant differences between the values of the parameters describing the disappearance of muon neutrinos and antineutrinos were observed. An analysis using an effective two-flavor neutrino oscillation model where the sine of the mixing angle is allowed to take non-physical values larger than 1 is also performed to check the consistency of our data with the three-flavor model. Our data were found to be consistent with a physical value for the mixing angle.

%% file: bodytext.tex

\section{Introduction}
We present an update of T2K's $\nu_{\mu}$ and $\bar{\nu}_{\mu}$ disappearance measurement from \cite{Abe:2017bay} with a larger statistical sample and significant analysis improvements. Data taken up until the end of 2018 are used. This is a beam exposure of $1.49 \times 10^{21}$ ($1.64 \times 10^{21}$) protons on target in neutrino (antineutrino) mode; an increase by a factor of 2.0 (2.2) over the previous result. While the same data were used for the result reported in \cite{Abe:2019vii}, the result reported here focuses on events containing $\nu_{\mu}$ and $\bar{\nu}_{\mu}$ candidates. These events are used to search for potential differences between neutrinos and antineutrinos and to test consistency with the PMNS oscillation model, by adding additional degrees of freedom to the oscillation probability formulae in the present analysis. These additional degrees of freedom are more straightforward to implement and interpret when studying muon-like events only.  

The mixing of the three flavors of neutrinos without sterile neutrinos or non-standard interactions is usually described with the PMNS formalism~\cite{Maki:1962mu,Pontecorvo:1967fh}. In this formalism the vacuum oscillation probability is determined by 6 parameters: three angles ($\theta_{12}$, $\theta_{13}$ and $\theta_{23}$), two mass squared splittings ($\Delta m^2_{21}$ and $\Delta m^2_{32}$, where $\Delta m^2_{ij}=m^2_{i}-m^2_{j}$) and a complex phase ($\delta_{CP}$). It is not known whether the smaller of the two mass splittings is between the two lightest states or the two heaviest states. These two cases are called normal and inverted ordering, respectively. $\nu_{\mu}$ disappearance is not sensitive to this ordering, so all results here assume the normal mass ordering.

In this model, which assumes CPT conservation, $\nu_{\mu}$ and $\bar{\nu}_{\mu}$ have identical survival probabilities for vacuum oscillations. At T2K's beam energy and baseline, the effect of the neutrinos propagating through matter on the muon neutrino survival probability is very small. Therefore, if the oscillation probabilities for neutrinos and antineutrinos differ by significantly more than expected, this could be interpreted as possible CPT violation and/or non-standard interactions \cite{PhysRevD.85.096005,Miranda_2015}.

In the three-flavor analysis shown here, the oscillation probabilities for $\nu_{\mu}$ and $\bar{\nu}_{\mu}$ are calculated using the standard PMNS formalism, but with independent parameters to describe $\bar{\nu}_{\mu}$ and $\nu_{\mu}$ oscillations, i.e. $\bar{\theta}_{23}\neq\theta_{23}$ and $\overline{\Delta m^2}_{32}\neq\Delta m^2_{32}$, where the barred parameters affect the antineutrino probabilities. As this data set does not constrain the other PMNS parameters, they are assumed to be the same for $\nu$ and $\bar{\nu}$.

Whilst it allows the $\nu_{\mu}$ and $\bar{\nu}_{\mu}$ parameters to take different values, this three-flavor analysis does not allow oscillation probability values not allowed by the PMNS formalism. To test consistency with the PMNS formalism we also present an analysis in which the oscillation probabiity is allowed to exceed the maximum possible PMNS value. In this analysis for computational simplicity we approximate the probability for muon neutrino disappearance using a `two-flavor' only oscillation formula with an effective mixing angle and mass splitting that takes into account the information we know about `three-flavor' mixing. $\sin^2\left(2\theta\right)$ is then allowed to take values exceeding 1, where $\theta$ is the effective neutrino mixing angle in this framework. This two-flavor approximation gives probabilities that agree to better than 0.5\% with the full PMNS calculation across T2K's neutrino energy range at the best-fit parameter values from T2K's joint muon and electron-like event analysis \cite{Abe:2019vii}.

\section{Experimental Apparatus}
T2K \cite{Abe:2011ks} searches for  neutrino oscillations in a long-baseline (295 km) neutrino beam sent from the Japan Proton Accelerator Research Complex (J-PARC) in Tokai, Japan to the Super-Kamiokande (SK) detector. SK \cite{Fukuda:2002uc,Abe:2013gga}, is situated $2.5^\circ$ off the axis of the beam, meaning that it is exposed to a relatively narrow energy width neutrino flux, peaked around the oscillation maximum 0.6 GeV. 

The neutrino beam generation starts with 30 GeV protons which strike a graphite target, producing hadrons, which are charge-selected and focused by three magnetic horns \cite{Sekiguchi:2015ghw}, and decay in a 96~m long decay volume producing predominantly muon neutrinos. Positively or negatively charged hadrons are selected using the polarity of the horns creating a beam dominated by neutrinos or antineutrinos, respectively.

A set of near detectors measures the unoscillated neutrino beam 280 m downstream of the interaction target. The INGRID \cite{Otani:2010zza} detector is an array of iron/scintillator sandwiches arranged in a cross pattern centered on the beam axis. INGRID measures the neutrino beam direction, stability and profile \cite{Suzuki:2014jyd}.

The off-axis ND280 detector has three magnetised time projection chamber (TPC) trackers \cite{Abgrall2011} and two fine-grained detectors (FGD1 made of CH, and FGD2 made of 52\% water 48\% CH by mass) \cite{Amaudruz2012}, surrounded by an electromagnetic calorimeter \cite{Allan:2013ofa}. A muon range detector \cite{Aoki:2012mf} is located inside the magnet yokes. The magnetized tracker measures the momentum and charge of particles. ND280 constrains the $\nu_\mu$ and $\overline{\nu}_{\mu}$ flux, the intrinsic $\nu_e$ and $\overline{\nu}_e$ contamination of the beam and the interaction cross sections of different neutrino reactions.

The far detector, SK \cite{Fukuda:2002uc,Abe:2013gga} is a 50 kt water Cherenkov detector, equipped with 11,129 inward facing 20-inch  photomultiplier tubes (PMTs) that image neutrino interactions in the pure water of the inner detector. SK also has 1,885 outward-facing 8-inch PMTs instrumenting the outer detector, used to veto events with interaction vertices outside the inner detector.

\section{Analysis Description}
The analysis presented here follows the same strategy as T2K's PMNS three-flavor joint fit to muon disappearance and electron appearance data in \cite{Abe:2019vii}. A model is constructed that gives predictions of the spectra at the near and far detectors. This model uses simulations of the neutrino flux, interaction cross sections and detector response and has variable parameters to account for both systematic and oscillation parameters. First a fit of this model is performed to the near-detector data to tune and constrain the neutrino flux and interaction cross-section uncertainties. The results of this fit are then propagated to the far detector as a multivariate normal distribution described by a covariance matrix and the best-fit values for each systematic parameter. The far-detector data are then fit to constrain the oscillation parameters. This section describes each part of the analysis focussing on changes from the analysis reported in \cite{Abe:2017bay}. Where not stated the same procedure as in \cite{Abe:2019vii} is used. Particularly, the beam flux prediction, neutrino interaction modeling, systematic uncertainties and near detector event selection are unchanged and the far-detector event selection used in this result is a subset of that in \cite{Abe:2019vii}.

\subsection{Beam flux prediction}
The T2K neutrino flux and energy spectrum prediction is discussed extensively in \cite{Abe:2013fp}. The modeling of hadronic interactions is constrained by thin target hadron production data, from the NA61/SHINE experiment at CERN \cite{Abgrall:2011ae,Abgrall:2011ts,Abgrall:2015hmv,Posiadala-Zezula:2017ivt,Zambelli:2017klm}. Before the ND280 analysis, the systematic uncertainties on the expected number of muon-like events after oscillations at SK due to the beam flux model are 8\% and 7.3\% for the $\nu_{\mu}$ and $\bar{\nu}_{\mu}$ beams, respectively. 

\subsection{Neutrino interaction models}
The $\nu_{\mu}$ and $\bar{\nu}_{\mu}$ oscillation probabilities are expected to be symmetric, but their interaction probabilities with matter are not. For example, the interaction cross section for a charged-current quasielastic (CCQE) $\nu_{\mu}$ interaction on oxygen, is about 4 times higher than that for $\bar{\nu}_{\mu}$.

We model neutrino interactions using the NEUT interaction generator \cite{neut}. The interaction cross-section model and uncertainties used in this result are the same as in \cite{Abe:2019vii}. This model is significantly improved compared to the previous version of this analysis \cite{Abe:2017bay}. The treatment of multinucleon so-called 2p2h interactions \cite{nieves,nievesExtension} has been updated, with new uncertainties accounting for different rates of this interaction for neutrinos and antineutrinos and for carbon and oxygen targets. We also allow the shape of the interaction cross section for 2p2h in energy-momentum transfer space to vary between that expected for a fully $\Delta$-exchange type interaction and that expected for a fully non-$\Delta$-exchange like interaction.

An uncertainty on the shielding of nucleons by the nucleus in CCQE interactions, modeled using the Nieves random phase approximation (RPA) method, has been added to the analysis \cite{nieves_rpa,nieves_rpa_erratum,nieves_rpa_uncertainty,rik_rpa}. The analysis also now accounts for mismodeling that could take place due to choosing an incorrect value for  the nucleon removal energy in the CCQE process. Finally, a fit to external data \cite{Abe:2018wpn,Stowell_2017} is now used to constrain our uncertainties on resonant single-pion production.

\subsection{Near detector event selection}

We define 14 samples of near-detector events, each targeting a particular part of our flux or cross-section model. All selected events must have a reconstructed charged muon present as the highest momentum track, as we are targeting charged-current (CC) neutrino interactions. In neutrino beam mode, the muon is required to be negatively charged to target neutrino interactions. The neutrino mode samples are separated by the number of pions reconstructed: 0, 1 positively charged pion and any other number of pions, giving samples enriched in CCQE, CC single pion and CC deep inelastic scattering interactions, respectively. 

In antineutrino beam mode there is one set of samples for positively charged muons and one set for negatively charged muons, allowing a separate constraint of the neutrino and antineutrino composition of the beam. This is important in antineutrino mode as the interaction cross section for neutrinos is larger than for antineutrinos. The antineutrino mode samples are separated by the number of reconstructed tracks matched between the TPC and FGD: 1 or more than 1, giving samples enriched in CCQE and CC non-QE interactions, respectively. In both beam modes, samples are further separated by which FGD their vertices are reconstructed in. As in \cite{Abe:2019vii}, the near-detector data set for antineutrino mode is 1.38 times larger than in \cite{Abe:2017bay}, while the neutrino mode data set is the same size.


\subsection{Far detector event selection}
The analyses presented here target muon-like events. SK is not able to distinguish neutrinos from antineutrinos at an event by event level as it cannot reconstruct the charge of the resulting muons. Hence, we form separate samples of events from neutrino and antineutrino beam mode to separately measure $\nu_{\mu}$ and $\bar{\nu}_{\mu}$ oscillations.

SK's vertex position, momentum, and particle identification (PID) are reconstructed from the Cherenkov rings produced by charged particles traversing the detector. PID is possible because muons scatter little due to their large mass and hence produce a clear ring pattern, while electrons produce electromagnetic showers resulting in Cherenkov rings with diffuse edges. The ring's opening angle also helps to distinguish between electrons and muons. The samples used here require exactly one muon-like Cherenkov ring and no other rings to be reconstructed and are referred to as 1R$\mu$. 

T2K's reconstruction algorithm \cite{Jiang_2019} fits the number of photons and timing information from each SK PMT, allowing better signal-background discrimination and a fiducial volume increase of $\sim$20\% over the previous algorithm used in \cite{Abe:2017bay}. Both 1R$\mu$ samples use the same selection criteria as in \cite{Abe:2019vii}. Table\,\ref{tab:NPred} shows the number of events predicted and observed for both 1R$\mu$ samples.



\begin{table}[]
    \centering
    \begin{tabular}{l|c|c}
        \hline
        Sample & Prediction & Data \\
        \hline
        $\nu$-mode 1R$\mu$ &  272.34 & 243\\
        $\bar\nu$-mode 1R$\mu$ & 139.47 & 140\\
        \hline
    \end{tabular}
    \caption{Number of events predicted using the best-fit oscillation parameter values from a previous T2K analysis \cite{Abe:2018wpn}, and the number of data events collected for both 1R$\mu$ samples.}
    \label{tab:NPred}
\end{table}

\subsection{Systematic uncertainties and oscillation analysis}
Our model includes systematic uncertainties from the neutrino flux prediction, the neutrino interaction cross-section model and detector effects. We constrain several of these uncertainties by fitting our model to ND280 near-detector data in bins of muon momentum and angle. This ND280 constrained model is then used as the prior in the fits to the far-detector data, where the SK muon-like samples are binned in the neutrino energy reconstructed using lepton momentum and angle assuming a CCQE interaction. Table \ref{tab:systs} shows the total systematic error in each 1R$\mu$ sample and a breakdown of the contributions from each uncertainty source. The near-detector fit introduces large anticorrelations between the parameters modeling the flux and cross-section uncertainties, so Table \ref{tab:systs} also lists the overall contribution to the uncertainty from the combination of flux and cross-section uncertainties.

The near-detector fit reduces the systematic error on the expected number of events in the neutrino (antineutrino) mode 1R$\mu$ sample from 15 (13)\% to 5.5 (4.4)\%.


\begin{table}[!htbb]
    \centering
    \begin{tabular}{l|c|c}
        \hline
        Error source & 1R$\mu$ $\nu$-mode & 1R$\mu$ $\bar{\nu}$-mode \\
        \hline
        Flux (constr. by ND280) & 4.3\% & 4.1\% \\
        Xsec (constr. by ND280) & 4.7\% & 4.0\%\\ 
        Xsec (all) & 5.6\% & 4.4\% \\
        \hline
        Flux + Xsec (constr. by ND280) & 3.3\% & 2.9\% \\
        Flux + Xsec (all) & 5.4\% & 3.2\% \\
        \hline
        SK detector effects+FSI+SI & 3.3\% & 2.9\% \\
        \hline
        \hline
        \textbf{Total} & 5.5\% & 4.4\% \\
        \hline
        \hline
    \end{tabular}
    \caption{Systematic uncertainty on the number of events in each of the 1R$\mu$ samples broken down by uncertainty source. Neutrino cross-section parameter uncertainties (denoted `xsec') are broken down by whether they are constrained by ND280 data or not. Uncertainties due to final state interactions (FSI) and secondary interactions (SI) are incorporated in the analysis by adding them to the SK detector effect uncertainty, so these are listed together.}
    \label{tab:systs}
\end{table}


In the three-flavor analysis, oscillation probabilities for all events are calculated using the full PMNS formulae \cite{Barger:1980tf}, with matter effects (crust density, $\rho=2.6\,\text{g}/\text{cm}^3$\, \cite{Hagiwara2011}). We allow the values of $\theta_{23}$ and $\Delta m^2_{32}$ used in the neutrino oscillation probability calculation to vary independently from those used for the antineutrino oscillation probability, in order to search for differences between neutrino and antineutrino oscillations.


In the two-flavor analysis, we use a modified version of the canonical two-flavor oscillation formula \cite{Nunokawa:2005nx}, where the disappearance probability for $\nu_{\mu}$ ($\bar{\nu}_{\mu}$) is given by:
\begin{equation*}
    P_{\nu_{\mu}\rightarrow\nu_{\mu}}\left( P_{\bar{\nu}_{\mu}\rightarrow\bar{\nu}_{\mu}}\right) \approx 1-\alpha(\bar{\alpha}) \sin^2\left(1.267\frac{\Delta m^2[\text{eV}^2] L[\text{km}]}{E[\text{GeV}]}\right)
\end{equation*}
where $\alpha$ plays the role of the well-known effective two flavor mixing angle, $\sin^22\theta$. $\alpha$ differs from $\sin^22\theta$ in that it is allowed to take values larger than 1. The effective two-flavor $\Delta m^2$ used here can be obtained from the three-flavor oscillation parameters using the following equation:
\begin{align*}
  \Delta m^2 = &\Delta m^2_{32} + \sin^2\theta_{12} \Delta m^2_{21} \\ 
  &+ \cos \delta_{CP} \sin \theta_{13} \sin 2\theta_{12} \tan \theta_{23}\Delta m^2_{21}.  
\end{align*}
We use independent oscillation parameters for neutrinos and antineutrinos, with $\alpha$ and $\Delta m^2$ affecting neutrinos, and $\bar{\alpha}$ and $\overline{\Delta m^2}$ affecting antineutrinos.

When $\alpha>1.0$, the $\nu_{\mu}$ ($\bar{\nu}_{\mu}$) survival probability is negative at some points in ($\Delta m^2$,$E_{\nu}$) parameter space. When weighting our Monte Carlo to produce predicted spectra for these points of parameter space this gives negative oscillation probability weights for some events. We allow these negative event weights, but we do not allow the total predicted number of events in any bin of our event samples to be negative, setting them instead to $10^{-6}$ where this occurs.


For both the two-flavor and three-flavor analyses, a joint maximum-likelihood fit to both 1R$\mu$ samples is performed. The likelihood used is a marginal likelihood where all parameters except the parameters of interest are marginalized over. 



The priors for the nuisance parameters are taken from the uncertainty model after the fit to ND280 data. Uniform priors are used in $\delta_{CP}$, $\Delta m^2_{32}$ and $\sin^2\theta_{23}$. $\theta_{12}$ and $\Delta m^2_{12}$ are fixed at their values from \cite{PDG}, due to their negligible effect on the $\nu_{\mu}$ survival probability. The prior on $\theta_{13}$ is taken from \cite{PDG}.



%

We build frequentist confidence intervals assuming the critical values for $\Delta \chi^2$ from a standard $\chi^2$ distribution. $\Delta \chi^2$ is defined as the difference between the minimum $\chi^2$ and the value for a given point in parameter space. 


\section{Results and discussion}
\begin{figure}
\includegraphics[width=\columnwidth]{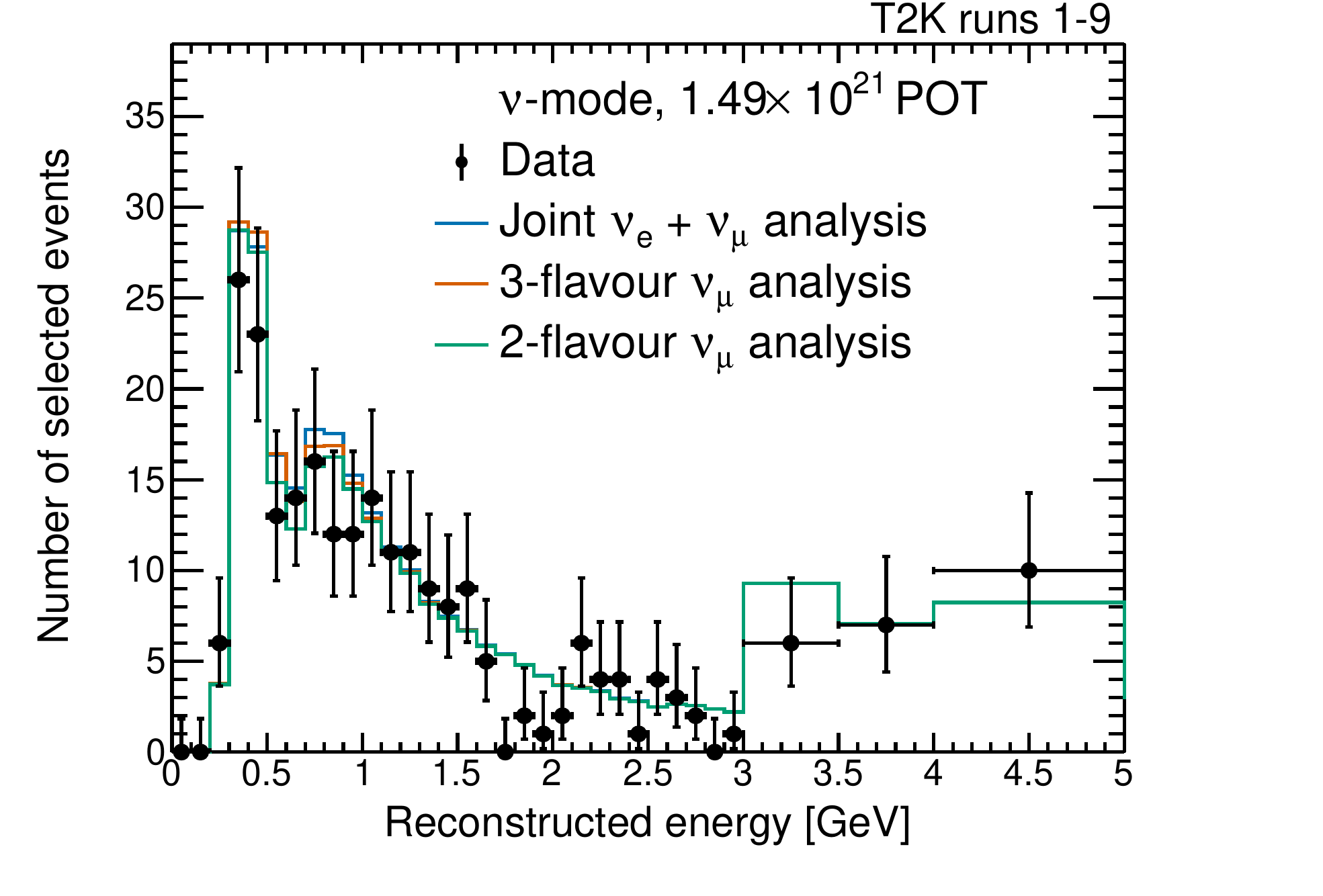} 

\includegraphics[width=\columnwidth]{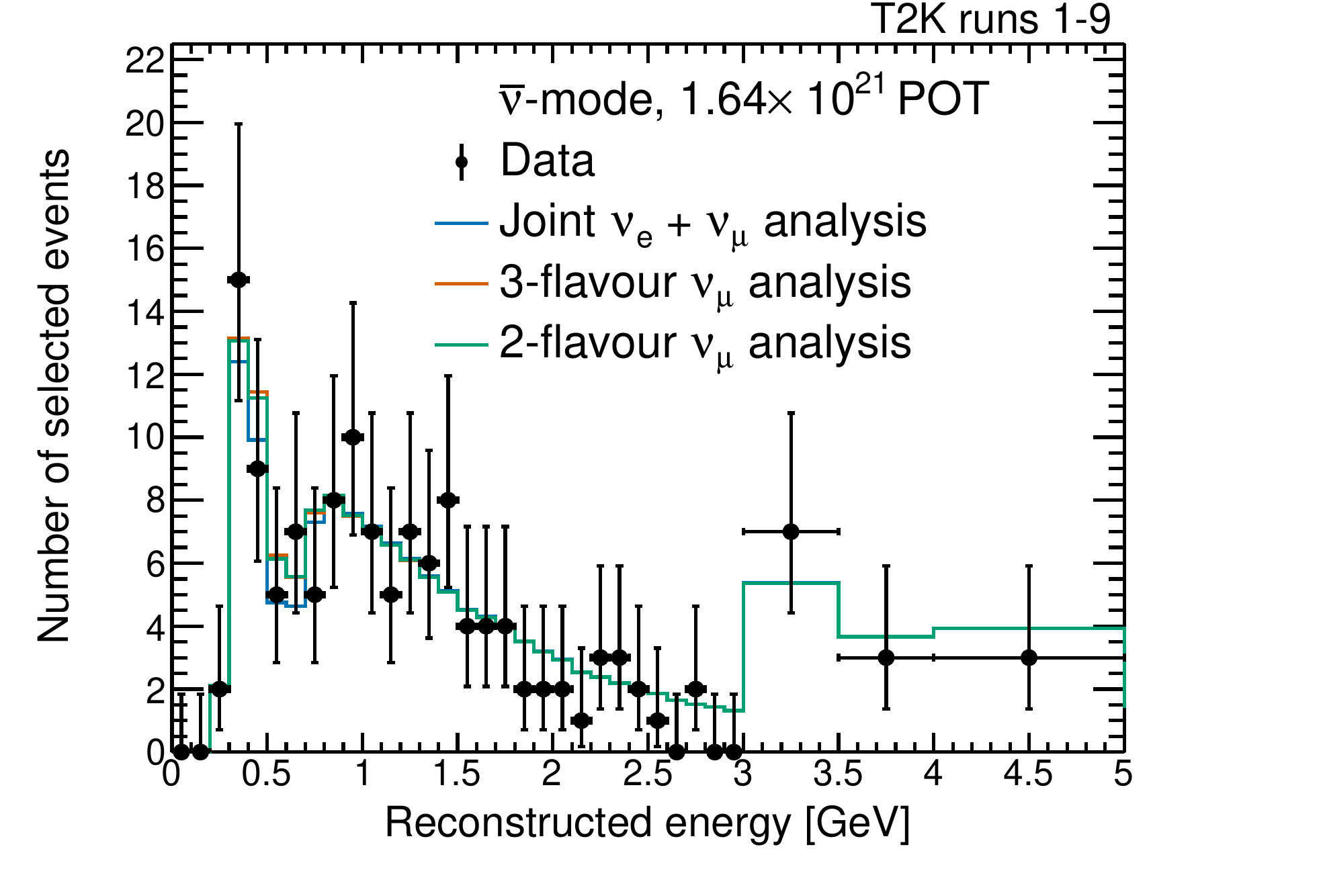}
\caption{Reconstructed energy spectra for the neutrino mode (top) and antineutrino mode (bottom) 1R$\mu$ samples. The lines show the predicted number of events under several oscillation hypotheses: `Joint $\nu_{e}/\nu_{\mu}$ analysis' uses the best-fit values from a joint fit of the PMNS model to electron-like and muon-like data \cite{Abe:2019vii}, `3-flavor $\nu_{\mu}$ analysis' uses the best fit from three-flavor fit reported here to the muon-like data, `2-flavor $\nu_{\mu}$ analysis' uses the best-fit value in the two flavor fit reported here to the muon-like data. The uncertainty on the data includes all predicted event rates for which the measured number of data events is less than a Poisson standard deviation from that prediction.}
\label{fig:2flavspectra}
\end{figure}

The reconstructed energy spectra of the $\nu_\mu$ and $\bar\nu_\mu$ events observed during neutrino and antineutrino running modes are shown in Fig.\,\ref{fig:2flavspectra}. All fits discussed below are to both 1R$\mu$ samples unless stated otherwise.

\subsection{Three-flavor analysis}
 For normal ordering, the best-fit values obtained for the parameters describing neutrino oscillations are \sinsqthetamu=$0.51^{+0.06}_{-0.07}$ and \Deltamu=$2.47^{+0.08}_{-0.09}\times10^{-3}~\text{eV}^2/\text{c}^4$, and those describing antineutrino oscillations are \sinsqthetamub=$0.43^{+0.21}_{-0.05}$ and \Deltamub=$2.50^{+0.18}_{-0.13}\times10^{-3}~\text{eV}^2/\text{c}^4$. The best-fit value and uncertainty on $\overline{\Delta m^2_{32}}$ obtained for normal ordering are equivalent to those that would be obtained on $\overline{\Delta m^2_{31}}$ for inverted ordering. 

\begin{figure}[!htb]
    \centering
    \includegraphics[width=0.99\linewidth]{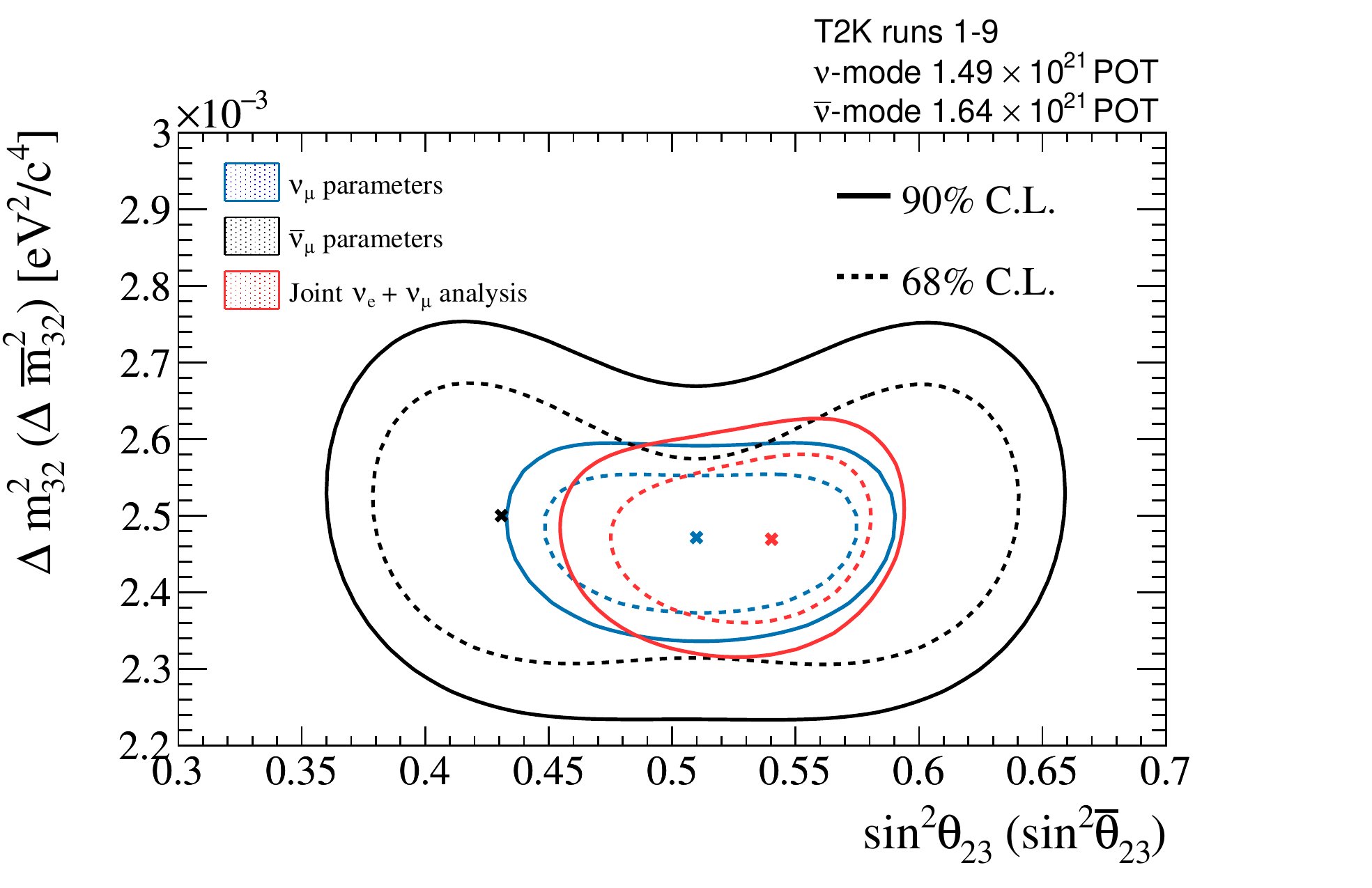}
    \caption{
        68\% and 90\% confidence intervals on \sinsqthetamu\, and \Deltamu\, (blue) and \sinsqthetamub\, and \Deltamub\ (black) from the three-flavor analysis described here. Also shown are equivalent intervals on $\sin^2\theta_{23}$ and $\Delta m^2_{32}$ (red) from a joint fit to muon-like and electron-like T2K data described in \cite{Abe:2019vii}.
    }
    \label{fig:Atm1RmuResults}
\end{figure}

Fig.\,\ref{fig:Atm1RmuResults} shows the confidence intervals on the oscillation parameters applying to $\nu_{\mu}$ overlaid on those for the parameters applying to $\bar{\nu}_{\mu}$. As the parameters for $\nu_{\mu}$ and $\bar{\nu}_{\mu}$ show no significant incompatibility, this analysis provides no indication of new physics. We also show the confidence interval for $\Delta m^2_{32}$ and $\sin^2\theta_{23}$ from the fit to electron-like and muon-like data in \cite{Abe:2019vii}. One can see by comparing these results that T2K's sensitivity to whether \sinsqthetamu\, is above or below 0.5 is driven by the electron-like samples, as the $\nu_{\mu}$ disappearance probability depends at leading order on the $\sin^2\left(2\theta_{23}\right)$.

\subsection{Two-flavor consistency check analysis}
\label{subs:Two-flavor consistency check analysis}
\begin{figure}
    \centering
   \includegraphics[width=\columnwidth]{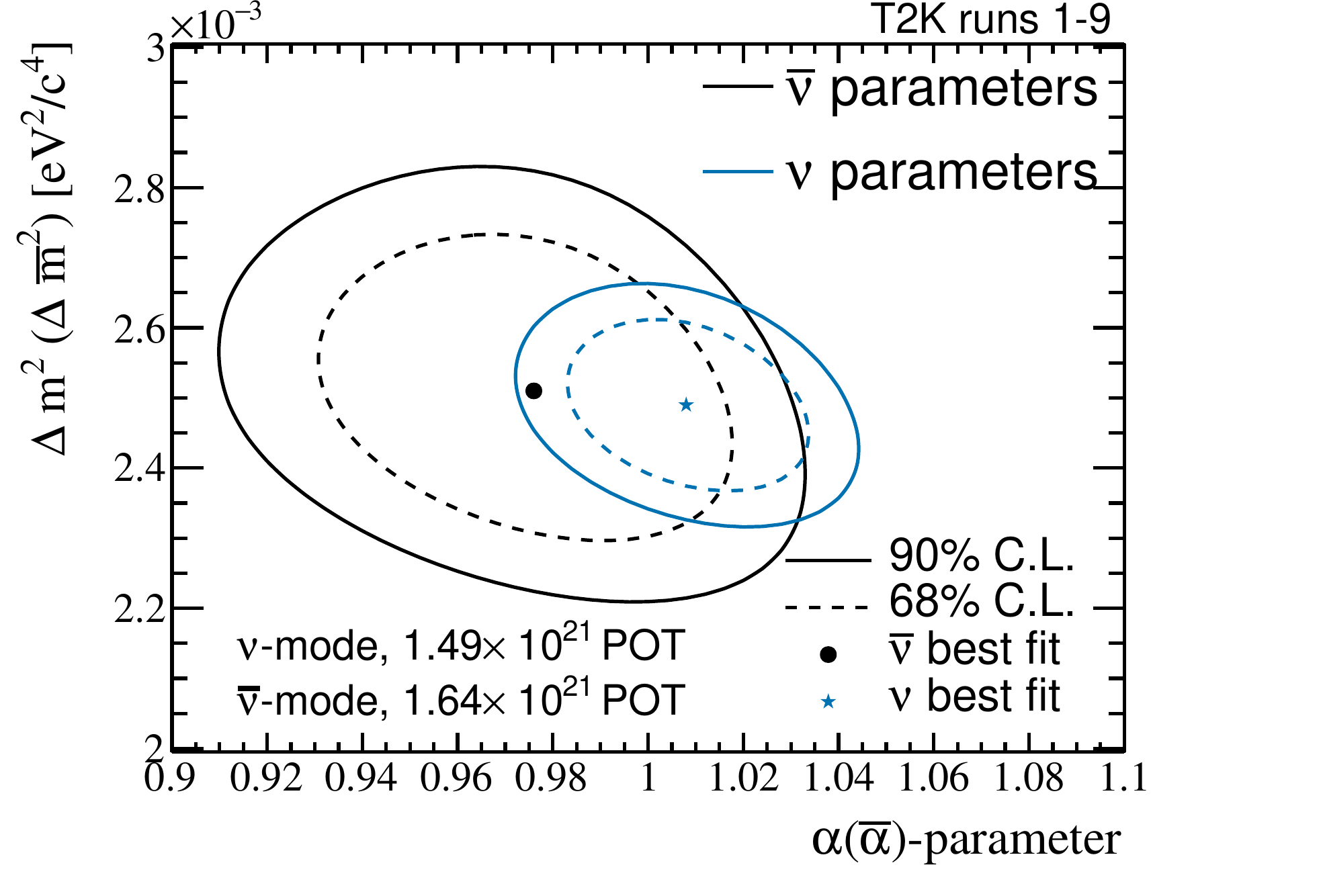}
    \caption{68$\%$ and 90$\%$ confidence intervals on the two-flavor analysis parameters affecting neutrinos (\Dmunull,$\alpha$), and antineutrinos (\Dmunullb,$\bar{\alpha}$).}
    \label{fig:2flavresult}
\end{figure}

  The best-fit values obtained on the effective two-flavor oscillation parameters are \Dmueff = $2.49^{+0.08}_{-0.08}$ $\text{eV}^2/c^4,$ $\alpha=1.008^{+0.017}_{-0.016}$, \Dmueffb = $2.51^{+0.15}_{-0.14}$\evmass,  $\bar{\alpha}=0.976^{+0.029}_{-0.029}$. Fig.~\ref{fig:2flavresult} shows the 68$\%$ and 90$\%$ confidence intervals for (\Dmueff, $\alpha$) and (\Dmueffb, $\bar{\alpha}$). Both the 1$\sigma$ confidence intervals include values of $\alpha (\bar{\alpha})\leq 1.0$, indicating no significant disagreement between data and standard physical PMNS neutrino oscillations. We also see good compatibility between the parameters affecting neutrinos and antineutrinos.

\subsection{Conclusions}
We have shown separate measurements of the oscillation parameters governing $\nu_{\mu}$ and $\bar{\nu}_{\mu}$ disappearance in long-baseline neutrino experiments using a significantly larger data sample and a much improved model of systematic uncertainties than those used in T2K's previous measurement of these parameters in \cite{Abe:2017bay}. We also show a consistency check between our data and the PMNS framework, where $\sin^2(2\theta)$ is allowed to take values larger than 1. In all analyses we find the neutrino and antineutrino oscillation parameters are compatible with each other, and that our data are compatible with the PMNS framework. The results from these fits improve upon the sensitivity of and are not in significant disagreement with previous similar results from the MINOS collaboration \cite{Adamson:2014vgd} (both show values of $\Delta m^2_{32}$ around $2.5$\evmass and $\theta_{23}$ consistent with maximal mixing).




%% file: acknowledgments.tex
\begin{acknowledgments}
We thank the J-PARC staff for superb accelerator performance. We thank the CERN NA61/SHINE Collaboration for providing valuable particle production data. We acknowledge the support of MEXT, Japan; NSERC (grant number SAPPJ-2014-00031), the NRC and CFI, Canada; the CEA and CNRS/IN2P3, France; the DFG, Germany; the INFN, Italy; the National Science Centre and Ministry of Science and Higher Education, Poland; the RSF (grant number 19-12-00325) and the Ministry of Science and Higher Education, Russia; MINECO and ERDF funds, Spain; the SNSF and SERI, Switzerland; the STFC, UK; and the DOE, USA. We also thank CERN for the UA1/NOMAD magnet, DESY for the HERA-B magnet mover system, NII for SINET5, the WestGrid and SciNet consortia in Compute Canada, and GridPP in the United Kingdom. In addition, participation of individual researchers and institutions has been further supported by funds from the ERC (FP7), “la Caixa” Foundation (ID 100010434, fellowship code LCF/BQ/IN17/11620050), the European Union’s Horizon 2020 Research and Innovation Programme under the Marie Sklodowska-Curie grant agreement numbers 713673 and 754496, and H2020 grant numbers RISE-GA822070-JENNIFER2 2020 and RISE-GA872549-SK2HK; the JSPS, Japan; the Royal Society, UK; French ANR grant number ANR-19-CE31-0001; and the DOE Early Career programme, USA.

\end{acknowledgments}